\documentstyle[12pt,procsla,epsfig]{article}
  
  \catcode`\@=11
  \long\def\@makefntext#1{
  \protect\noindent \hbox to 3.2pt {\hskip-.9pt  
  $^{{\ninerm\@thefnmark}}$\hfil}#1\hfill}		
  
  \def\@makefnmark{\hbox to 0pt{$^{\@thefnmark}$\hss}}  
  	
  \def\ps@myheadings{\let\@mkboth\@gobbletwo
  \def\@oddhead{\hbox{}
  \rightmark\hfil\ninerm\thepage}   
  \def\@oddfoot{}\def\@evenhead{\ninerm\thepage\hfil
  \leftmark\hbox{}}\def\@evenfoot{}
  \def\sectionmark##1{}\def\subsectionmark##1{}}
  
\begin{document}
  
  \centerline{\normalsize\bf WHAT MIGHT WE LEARN FROM A FUTURE SUPERNOVA}
  \baselineskip=16pt
  \centerline{\normalsize\bf NEUTRINO SIGNAL?\footnote
{Invited talk at the $8^{th}$ Int. Workshop on ``Neutrino Telescopes'',
Venice, February 23-26, 1999.}}
  
  \vspace*{0.6cm}
  \centerline{\footnotesize PETR VOGEL}
  \baselineskip=13pt
  \centerline{\footnotesize\it Physics Department, Caltech }
  \baselineskip=12pt
  \centerline{\footnotesize\it Pasadena, CA 91125, USA}
  \centerline{\footnotesize E-mail: vogel@lamppost.caltech.edu}
  \vspace*{0.3cm}
  
  \vspace*{0.9cm}
  \abstracts{  Neutrinos from a future Galactic supernova will be detected 
by several large underground detectors, in particular by
SuperKamiokande (SK)  and 
the Sudbury Neutrino Observatory (SNO). 
If, as expected, the $\nu_{\mu}$ and  $\nu_{\tau}$
neutrinos have somewhat higher energy on average that
the electron neutrinos, they will dominate the neutral current 
response. The ways to separate the neutral and charged
current signals will be discussed, and the best strategy to
measure the possible time delay of the neutral current events
will be outlined. Given the expected count rates, one will
be able to measure in this way the  $\nu_{\tau}$ mass 
down to about 30 eV in SNO and to 50 eV in SuperKamiokande.
   Another application to be discussed is the supernova localization
by the neutrino signal, prior to or independently of the electromagnetic
signal. The accuracy with which this can be accomplished 
using the angular distributions of the reactions will be estimated.
With two or more detectors one can, in principle, attempt
triangulation based on the arrival time of the neutrinos.
It will be argued that for realistic parameters this method
will be very difficult and likely leads only to crude localization.}
   
  \vspace*{0.6cm}
  \normalsize\baselineskip=15pt
  \setcounter{footnote}{0}
  \renewcommand{\thefootnote}{\alph{footnote}}
  \section{Introduction}

When the core of a large star ($M \ge 8 M_{\odot}$) runs out of
nuclear fuel, it collapses and forms a proto-neutron star. The total energy
released in the collapse, i.e., the gravitational binding energy of
the core ($E_B \sim G_N M_ {\odot}^2/R$ with $R \sim$ 10 km), is about
$3 \times 10^{53}$ ergs; $\sim$ 99\% of that is carried away by
neutrinos and antineutrinos, the particles with the longest mean free
path.  It is believed that neutrinos of all three flavors
are emitted with approximately equal luminosities
over a timescale of several seconds.

Those flavors which interact the most with the matter will decouple at
the largest radius and thus the lowest temperature.  The $\nu_\mu$ and
$\nu_\tau$ neutrinos and their antiparticles 
have only neutral-current interactions with
the matter, and therefore leave with the highest temperature, about 8
MeV (or $\langle E \rangle \simeq$ 25 MeV).  The $\bar{\nu}_e$ and
$\nu_e$ neutrinos have also charged-current interactions, and so leave
with lower temperatures, about 5 MeV ($\langle E \rangle \simeq$ 16
MeV) and 3.5 MeV ($\langle E \rangle \simeq$ 11 MeV), respectively.
The $\nu_e$ temperature is lower because the material is neutron-rich
and thus the $\nu_e$ interact more than the $\bar{\nu}_e$.  The
observation of supernova 
$\nu_\mu$ and $\nu_\tau$ neutrinos and their antiparticles
would allow the details of the picture above to be tested.  

In this talk I concentrate on two aspects of the neutrino signal:
\begin{itemize}
\item The possibility of measuring or constraining the mass of the
$\nu_\tau$ and/or  $\nu_\mu$.
\item The possibility of locating the supernova by its neutrino signal,
independently  of or prior to the optical observation.
\end{itemize}
The physics of these task is straightforward, but there are complications 
due to:  
\begin{itemize}
\item  The finite statistics of the neutrino signal.
\item The finite time duration of the signal.
\end{itemize}
The details of the work reported here can be found in the joint
work with John Beacom of Caltech \cite{SK,SNO,point}.
One can find a much more complete list 
of the relevant earlier references there.

Numerical supernova models suggest that the neutrino luminosity rises
quickly over a time of order 0.1 s, and then falls over a time of
order several seconds.  The rise is so fast
that the details of its shape are largely irrelevant
for our task.  We model
the luminosity fall by an
exponential  with time constant $\tau$ = 3 s.  The luminosity
then has a width of about 10 s, consistent with the SN 1987A
observations.  Later, I will show how the conclusions depend
on the luminosity decay time constant $\tau$.

\section{Neutrino mass determination}

The requirement that neutrinos do not overclose the universe gives a
bound for the sum of masses of stable neutrinos (see e.g., \cite{Raffelt}):
\begin{equation}
\sum_{i=1}^3 m_{\nu_i} \le 100 {\rm\ eV}\,.
\label{eq:cosmo}
\end{equation}
However, laboratory kinematic tests of neutrino mass currently give limits
for the masses compatible with the above cosmological bound only for
the electron neutrino, $m_{\bar{\nu}_e} \le 5$ eV \cite{Belesev}.
For the $\nu_\mu$ and $\nu_\tau$ they far
exceed the cosmological bound: $m_{\nu_\mu} < 170$ keV\cite{RPP}, and
$m_{\nu_\tau} < 18$ MeV  \cite{RPP}.  It is very unlikely that 
these mass limits can improve by the necessary orders
of magnitude any time soon.

When neutrinos are emitted by a supernova,  even a tiny
mass will make the velocity  less than for a massless particle,
and will cause a
measurable delay in the arrival time.  A neutrino with a mass $m$ (in
eV) and energy $E$ (in MeV) will experience an energy-dependent delay
(in s) relative to a massless neutrino in traveling over a distance D
(in 10 kpc) of
\begin{equation}
\Delta t(E) = 0.515 \left(\frac{m}{E}\right)^2 D\,.
\label{eq:delay}
\end{equation}
For a supernova at 10 kpc distance (approximately at the 
center of the galaxy), the delay for $\nu_e$ and $\bar{\nu}_e$ 
will be negligible, and their signal can
be used as a reference clock. The $\nu_\mu$ and $\nu_\tau$ neutrinos and
their antiparticles will interact only by the neutral current. Thus, 
in order to determine the $\nu_\tau$ and/or $\nu_\mu$ mass,  we should find
ways of separating the neutral and charged current signals, 
and of determining the possible time delay of the former 
with respect to the latter.

There are three neutral current reactions that give rise to 
potentially measurable signals:
a) neutrino-electron scattering (we will show below 
that it is difficult to separate
the charged and neutral current events in that case), 
b) neutral current excitation
of $^{16}$O nuclei in water, followed by the $\gamma$ 
emission as suggested in \cite{LVK}, 
and c) the neutral current deuteron disintegration (relevant for SNO).
In Table \ref{tab:rate} I show the corresponding numbers of events 
(see \cite{SK,SNO} for details how the table
was made and refs. \cite{SKD,SNOD} for 
description of the detectors) for the individual
reactions, calculated for the ``standard'' supernova defined above.

\begin{table}[h]
\caption{Calculated numbers of events expected in SK and SNO.
In SNO events in 1 kton of D$_2$O and  in 1.4 kton of H$_2$O are added. 
By $\nu_x$ we denote the combined effect of  $\nu_\mu$ and $\nu_\tau$,
each accounts for half of the events.
In all except the top row, the events caused by $\nu$ and $\bar{\nu}$
are added.}
\label{tab:rate}
\vspace{5 mm}
\begin{center}
\small
\begin{tabular}{|l|l|l|}
\hline
reaction & events in SK & events in SNO \\
\hline\hline
$\bar{\nu}_e + p \rightarrow e^+ + n$ & 8300  & 365 \\
\hline
$\nu_e + d \rightarrow e^- + p + p $ & -  & 160 \\
$\bar{\nu}_e + d \rightarrow e^+ + n + n $ &  &  \\
\hline
$\nu_x + d \rightarrow \nu_x + n + p $ & -  & 400 \\
\hline
$\nu_x+ ^{16}{\rm O} \rightarrow \nu_x + \gamma + X$ & 710  & 50 \\
\hline
$\nu_x+ ^{16}{\rm O} \rightarrow \nu_x + n + ^{15}{\rm O} $ & -  & 15 \\
\hline
$\nu_e + e^- \rightarrow \nu_e + e^-$ & 200  & 15 \\
\hline
$\nu_x + e^- \rightarrow \nu_x + e^-$ & 120 & 10 \\
\hline \hline
\end{tabular}
\end{center}
\end{table}

Given the assumed known time dependence of the supernova luminosity $L(t)$,
and assuming that all flavors develop in time the same way and keep their
temperatures constant, the arrival time of massive neutrinos is then described
by $L(t - \Delta t(E_{\nu}))$. Since in the neutral current scattering one cannot
determine the incoming neutrino energy $E_{\nu}$, we have at our disposal
only the {\it time distribution} of the events
\begin{equation}
\frac{dN}{dt} = C
\int dE_{\nu} f(E_{\nu}) \sigma(E_{\nu}) L(t - \Delta t(E_{\nu})) \,,
\label{eq:rate}
\end{equation}
where
$C$ is a constant proportional to $1/(D^2 \times \langle E_{\nu} \rangle)$
and $f(E_{\nu})$ is the thermal neutrino spectrum.

The only way one can decide whether there is 
a time delay or not is to compare the
neutral current time distribution  (called ``Signal'') 
with the time distribution
of the charged current events (called ``Reference''). Since 
$\nu_\tau$ and $\nu_{\mu}$ neutrinos and their antiparticles have higher
energies than  $\nu_e$ and $\bar{\nu}_e$, the neutral current events will
contain a substantial fraction of possibly delayed events, while the charged
current events will have no delay.

It turns out, see \cite{SK}, that the most efficient way 
to accomplish this is also the
simplest one, i.e., to use the diffence in the 
{\it mean arrival time}:
\begin{equation}
\langle t \rangle_S = \sum_k t_k /N_S \,, 
~ \langle t \rangle_R = \sum_k t_k /N_R \,,
\end{equation}
where $N_S (N_R)$ is the total number 
of the Signal (Reference) events, and $t_k$
are the arrival times of the individual events.
The signature of neutrino mass is then the inequality
\begin{equation} 
 \langle t \rangle_S > \langle t \rangle_R \,,
 \end{equation}
valid with significance beyond statistical fluctuations.

\begin{figure}[h]
\vspace*{13pt}
\begin{center}
\mbox{\epsfig{figure=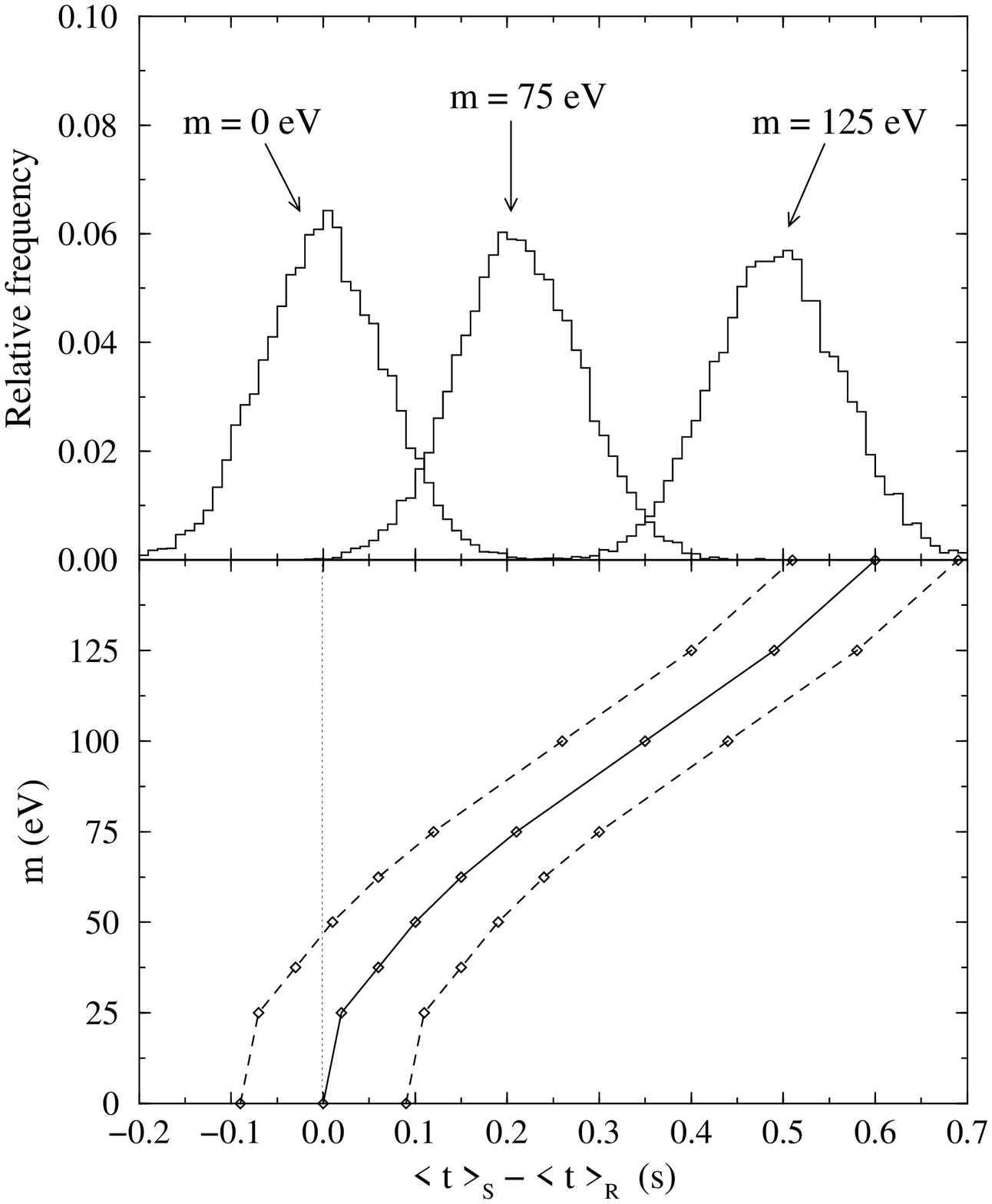,width=8cm}}
\vspace*{13pt}
\fcaption{The results of the $\langle t \rangle$ analysis for a massive
$\nu_\tau$ in SK using the
$\gamma$ following $^{16}$O excitation.  
In the upper panel, the relative frequencies of various
$\langle t \rangle_S - \langle t \rangle_R$ values are shown for a few
example masses.  In the lower panel, the range of masses corresponding
to a given $\langle t \rangle_S - \langle t \rangle_R$ is shown.  The
solid line is the 50\% confidence level, and the upper and lower
dashed lines are the 10\% and 90\% confidence levels, respectively.}
\label{fig:mom}
\end{center}
\end{figure}

The analysis below is based on the assumption that only one of the
neutrino flavors is massive, say $\nu_{\tau}$, and the other one,
$\nu_{\mu}$ in this case, is either massless or has so much smaller mass
that the corresponding time delay is negligible. The ``Signal'' then consist
of part that is delayed and another part that is not because it is either 
caused by the massless $\nu_{\mu}$ or belongs to background that cannot
be separated from the signal since it has the same energy and angle, etc.
With these assumptions, the neutrino-electron scattering signal in SK
will contain 60 delayed events and 700 background events, since a rather large
number of the charged current $\bar{\nu}_e + p \rightarrow e^+ + n$ events
will be present in the forward cone. For the $\gamma$ signal from the
$^{16}$O excitation, the ratio delayed/background is a more favorable 355/885.
And in SNO the neutral current deuteron disintegration, with a single
neutron and no charged lepton, is characterized by the ratio 219/316. 

The last two ratios above also show that events which look like
the neutral current (i.e., the true neutral current plus background with 
similar characteristics) are
dominated by the response to $\nu_{\tau}$ and $\nu_{\mu}$ neutrinos.
Indeed, since the cross sections are the same for these two flavors,
one can simply multiply the numerators by a factor of two and make
the corresponding adjustment in the denominator, to obtain
the fraction of events caused by the $\nu_{\tau}$ and $\nu_{\mu}$ neutrinos.
So, for the $\gamma$ signal from the $^{16}$O excitation, that contribution
is about 57\%, and for the deuteron disintegration it is about 82\%. On the other
hand, for the neutrino-electron scattering it is only about 16\%.
By measuring the total number of the Signal events one can determine
the temperatures of the  $\nu_{\tau}$ and $\nu_{\mu}$ neutrinos
(see \cite{SK,SNO}) with reasonable accuracy.

To judge the statistical significance of the delay, we used the Monte Carlo
simulation of a large number of supernovae for each mass value. We then
histogram the differences 
$ \langle t \rangle_S - \langle t \rangle_R$, and find
the 10\%, 50\%, and 90\% confidence levels. Representative cases are
plotted in Fig. 1 for the $\gamma$ from $^{16}$O
excitation in SK. In the upper panel one can see that if, 
e.g., $m_{\nu}$ = 75 eV 
the most probable difference in average arrival times is about 0.2 s, and the
10 - 90 \% CL band is 0.1 - 0.3 s, clearly separated from the massless case.
In fact, the smallest recognizable mass is about 50 eV.
In SNO, using the deuteron disintegration, 
the smallest recognizable mass is even
smaller, about 30 eV.

How does the mass sensitivity depend on the assumptions we made above?
Much of the analysis can be made analytically. We shall concentrate on
the dependence on the assumed temperatures $T$, distance $D$, and the
constant which characterizes the time duration 
of the neutrino signal, $\tau$. The time delay
and its error depend on
\begin{equation}
\langle t \rangle_S - \langle t \rangle_R  \sim (m/T)^2 D ~;~
\delta( \langle t \rangle_S - \langle t \rangle_R) \sim \tau D/\sqrt{T} ~.
\end{equation}
Since the significance of the result, and thus the smallest recognizable mass
$m_{lim}$,
is the ratio of these two quantities, we conclude that this neutrino mass limit
is, remarkably, independent on the distance $D$, and
\begin{equation}
m_{lim} \sim \sqrt{\tau}  T^{3/4} ~.
\end{equation}
Clearly, the shorter the duration of the neutrino pulse 
(i.e., smaller $\tau$), the
better the ability to determine the neutrino mass. 
(We verified that detailed numerical simulation closely follows the
$\sqrt{\tau}$ scaling above)
Also, naturally, if e.g. the
$\nu_{\mu}$ and $\nu_{\tau}$ masses are close to each other, the mass limit
is improved, roughly by $\sqrt{2}$.

\section{ Supernova localization with neutrinos}

A future core-collapse supernova in our Galaxy will be detected by several
neutrino detectors around the world.  The neutrinos escape from the
supernova core over several seconds from the time of collapse, unlike
the electromagnetic radiation, emitted from the envelope, which is
delayed by a time of order hours.  In addition, the electromagnetic
radiation can be obscured by dust in the intervening interstellar
space.  The question therefore arises whether a supernova can be
located by its neutrinos alone.  The early warning of a supernova and
its location might allow greatly improved astronomical observations.

There are two types of techniques to locate a supernova by its
neutrinos.  The first one is based on angular
distributions of the neutrino reaction products, which can be
correlated with the neutrino direction.  In this case, a single
experiment can independently announce a direction and its error.
However, to suppress false alarms  one
can demand coincidence with other experiments.
The second method of supernova
location is based on triangulation using two or more widely-separated
detectors.  This technique would require significant and immediate
data sharing among the different experiments. 
The theme of this section (for more details and more
complete reference list, see \cite{point}) 
is a careful and realistic assessment of
this question, taking into account the statistical significance of the
various neutrino signals.

\subsection{ Reactions with angular dependence}

Neutrino-electron scattering
occurs for all flavors of neutrinos and antineutrinos, and is detected
by observing the recoil electrons with kinetic energy $T$ .
The scattering angle is
dictated by the kinematics and is given by
\begin{equation}
\cos\alpha = \frac{E_{\nu} + m_e}{E_{\nu}} 
\left( \frac{ T}{T + 2m_e} \right)^{1/2}\,.
\end{equation}
With threshold of about $T_{\rm min} = 5$ MeV, the recoil electrons will
be sharply forward scattered, i.e., pointing away from the supernova,
with the combined
average $\langle \cos \alpha \rangle$ = 0.98, corresponding to about
$11^{\circ}$.  However, multiple scattering 
will smear the \v{C}erenkov cone,
resulting in a one-sigma width of $\sim 25^\circ$.
In order to evaluate the pointing ability  of this signal
we have to take into account 
the finite statistics, the two-dimensional form of the resulting distribution,
and the presence of the unavoidable background. The background worsens
the pointing ability from the simple expectation by a factor
\cite{point}  $C(R)$
\begin{equation}
\delta x = \frac{\sigma}{\sqrt{N_S}} \times C(R); 
~{\rm with} ~ C(R) \approx \sqrt{1 + 4 R} ~,
\end{equation}
where $R$ is the ratio (at the peak) 
of the flat background and the signal with $N_S$ events.
For SK and SNO the background reduction factor is $C(R) \simeq 2 - 3$, 
and with our standard supernova parameters
we find that the one-sigma error based 
on the neutrino-electron scattering will
be about $5^{\circ}$ in SK and about $20^{\circ}$ in SNO. 
This is by far the
most accurate pointing ability at our disposal.

The reaction with the most events is $\bar{\nu}_e + p
\rightarrow e^+ + n$, with $\simeq 10^4$ events expected in SK, and
$\simeq 400$ events expected in the light water of SNO. 
In \v{C}erenkov detectors one can determine the direction
of the positrons, whose angular distributions 
with respect to the direction of the neutrino beam is of the form
\begin{equation}
\frac{d N}{d \cos\alpha} = \frac{N}{2} \left(1 + a \cos\alpha \right)\,.
\label{eq:acos}
\end{equation}
It is relatively easy to show that the error
in the pointing ability for $N$ observed events in this case is given by
\begin{equation}
\delta (\cos\alpha) =
\frac{2}{|a|} \frac{1}{\sqrt{N}}\,.
\end{equation}
Since, in general  $a = a(E_\nu)$, we have to investigate further
the neutrino energy dependence of this coefficient, and perform 
the necessary energy averaging. 

\begin{figure}[h]
\vspace*{13pt}
\begin{center}
\mbox{\epsfig{figure=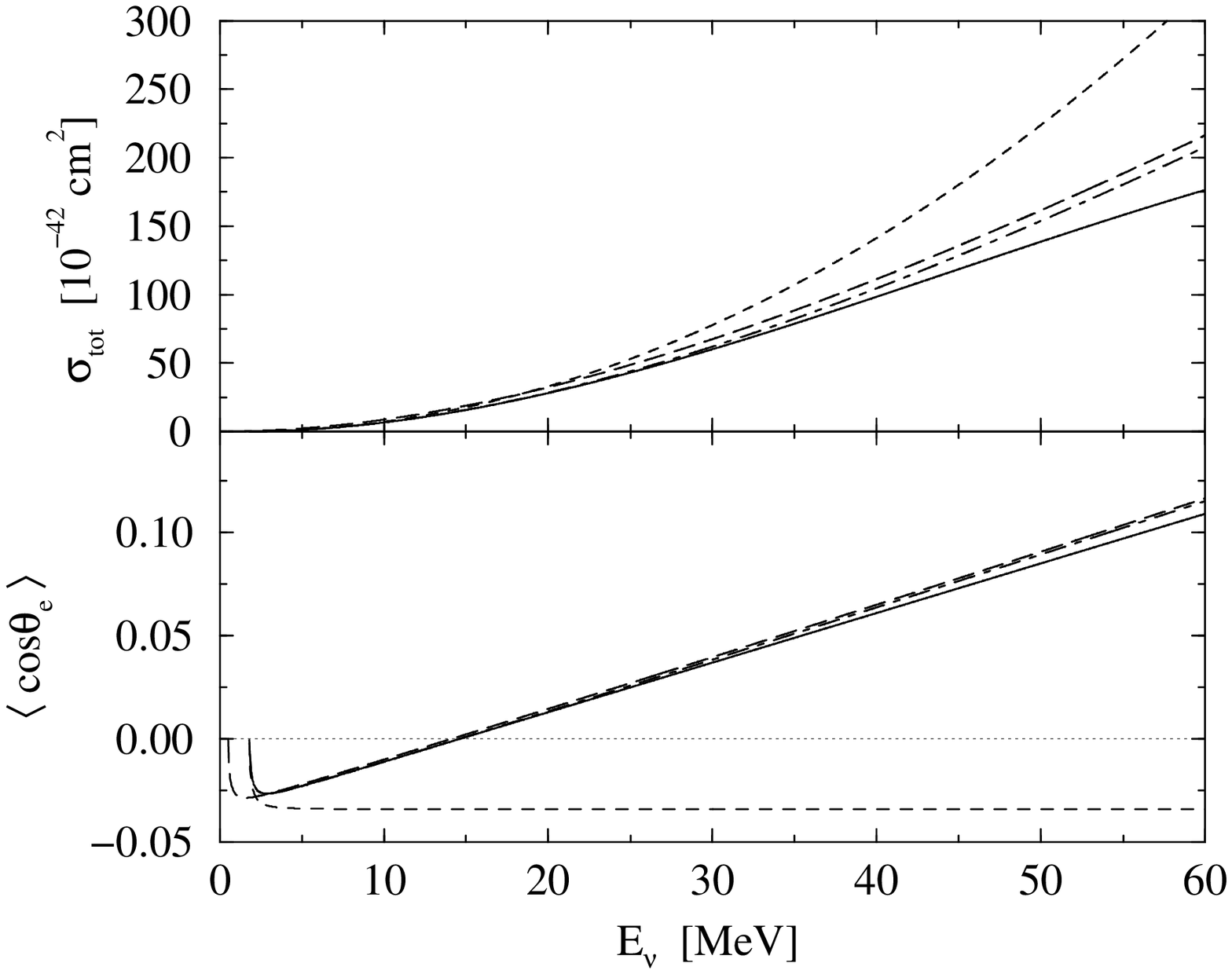,width=8cm}}
\vspace*{13pt}
\fcaption{Upper panel: total cross section for $\bar{\nu}_e + p
\rightarrow e^+ + n$; bottom panel: $\langle \cos\theta \rangle$ for
the same reaction; both as a function of the antineutrino energy.  The
solid line is the ${\cal O}(1/M)$ result and the short-dashed line is
the ${\cal O}(1)$ result.  The long-dashed line is the result of
Eq.(3.18) of Ref.~\protect\cite{LS}, and the dot-dashed line contains our
threshold modifications to the same.}
\label{fig:cos}
\end{center}
\end{figure}

In the limit where the nucleon mass $M$ is taken to be infinite, i.e.,
zeroth order in $1/M$ (${\cal O}(1)$), 
the asymmetry coefficient $a$ is independent
of $E_\nu$ and  is given simply by the competition of
the non-spin-flip (Fermi) and spin-flip (Gamow-Teller) contributions,
and is
\begin{equation}
a^{(0)} = \frac{f^2 - g^2}{f^2 + 3g^2} \simeq -0.10 ~~;~~
f = 1, g = 1.26 ~,
\label{eq:a0}
\end{equation}
and thus the angular distribution of the positrons is weakly backward.

However, $a(E_\nu)$ is substantially modified  when weak magnetism 
and recoil corrections of ${\cal O}(1/M)$ are included. It turns out \cite{angul}
that the inclusion to this order gives a very accurate formula for 
$\langle \cos \theta \rangle$. This quantity and the total
cross section are shown in Fig. \ref{fig:cos},
evaluated in various approximations (see \cite{angul}).
At high energies, the formula (3.18) of \cite{LS}, valid to all orders
in $1/M$, but neglecting the threshold effects, is applicable. One can see in 
Fig. \ref{fig:cos} that the dot-dashed line smoothly interpolates between the
correct low energy and high energy behaviour.

As far as the pointing ability of the 
$\bar{\nu}_e + p \rightarrow e^+ + n$ reaction is concerned,
due to the rather small angular asymmetry (and its energy
dependence) we estimate that the uncertainty 
$\delta (\cos\alpha) \simeq 0.2$ even for the high statistics
detector like SK. Nevertheless, it would be important and useful
to use this additional information constraining the supernova direction.

\subsection{ Triangulation}

For two detectors separated by a distance $d$, there will be a delay
between the arrival times of the neutrino pulse.  The magnitude of the
delay $\Delta t$ depends upon the angle $\theta$ between the supernova
direction and the axis connecting the two detectors.  Given a measured
time delay $\Delta t$, the unknown angle $\theta$ and its error
are then:
\begin{equation}
\cos\theta = \frac{\Delta t}{d} ~~;~~ 
\delta(\cos\theta) = \frac{\delta(\Delta t)}{d}\,.
\end{equation}
Thus two detectors define a cone along their axis with opening
$\cos\theta$ and thickness $2 \times \delta(\cos\theta)$ in which the
supernova can lie.  Obviously, in order to have a reasonable pointing
accuracy from triangulation, one will need $\delta(\Delta t) \ll d$.
(The Earth diameter is $d \approx 40$ ms.)
Following \cite{point} I discuss whether an appropriate time delay can be
defined, and what its error would likely be.
Basically, the question can be reduced to the following problem in statistics:
given $N$ events of duration $\tau$, is the uncertainty  $\delta(\Delta t)$
equal to $\tau/N$ (i.e., the interval between events) or the much larger
$\tau/\sqrt{N}$? 

The answer, of obvious practical significance, requires a degree of subtlety.
Let us model, as before, the time dependence of the neutrino pulse (i.e. the
supernova luminosity $L(t)$) by two exponentials, the increasing sharp rise
with time constant $\tau_1$ and the slow decay with the time constant $\tau_2$
($\tau_1 \ll \tau_2$). Now take the limit (unrealistic) of zero risetime
($\tau_1 \rightarrow 0$). Then, in fact, the first answer is applicable, i.e.,
$\delta(\Delta t) \rightarrow \tau_2/N$. One would then simply
determine the arrival time of the first event in each detector,
and the triangulation would be feasible, though still not very accurate.

But any finite leading edge, or background, would invalidate this picture. 
Moreover, we know that the leading edge has a finite duration related
to the shock propagation time in the supernova. The best strategy then
is to try to determine the rather sharp point of maximum rate $t_0$.
The error in its determination depends on the duration of the leading
edge $\tau_1$, which can be measured in the largest detector, and
on the number of events $N_1$ in the leading edge for the given detector,
\begin{equation}
(\delta t_0)_{min} ~\approx ~ \frac{\tau_1}{\sqrt{N_1}} ~.
\end{equation} 
At the same time, the number of events $N_1$ in the leading edge
depends somewhat indirectly on the total duration of the pulse,
since $N_1 \approx N \tau_1/\tau_2$. For the existing detectors,
this leads to rather large uncertainty, $\delta(\cos\theta) \approx 0.5$.
Nevertheless, if there will be several  large detectors
available in not too distant future, triangulation
would offer another handle to the supernova localization, besides
the obvious benefit of the false alarm elimination by the
coincidence requirement.

\section{Conclusions}

In this talk, which is based on the results of Refs. \cite{SK,SNO,point,angul},
I have shown that:
\begin{itemize}
\item The supernova signal caused by $\nu_{\tau} + \nu_{\mu}$
and their antiparticles can be isolated.
\item By measuring the average arrival time difference of the neutral
and charged current events, one will be able to (conservatively) determine
the upper limit for $m_{\nu_{\tau}}$ of 30-50 eV, representing
an improvement by $10^6$ when compared to the existing limits.
\item Neutrino electron scattering can be used for pointing with
accuracy of about $5^{\circ}$.
\item  $\bar{\nu}_e + p \rightarrow e^+ + n$ can be also used
for crude pointing, provided the correct differential cross section
is used. (Remembering that the naive formula suggests that the
positron are slightly backward while in reality they should be
slightly forward.) 
\item Triangulation appears to be difficult if the supernova signal
is going to last more than one second. But it would be useful if
more than two detectors (and even better if they are going to be large)
will participate in the warning network.
\end{itemize}

 \section{Acknowledgements}
The collaboration with John Beacom is gratefully acknowledged.
This work was supported in part by the US Department of Energy under
Grant No. DE-FG03-88ER-40397.

  \end{document}